# The Effect of Base-Pairing on the Shape Resonances of Nucleobases

Jishnu Narayanan S J,[1] Divya Tripathi,[1] Idan Haritan,[2] Amitava Adhikary,[3] Bhawana Pandey,[4] and Achintya Kumar Dutta[1,5]*

1. Department of Chemistry, Indian Institute of Technology Bombay, Powai, Mumbai 400076, India.

2. Alexander Kofkin Faculty of Engineering, Bar-Ilan University, Ramat Gan, 5290002 Israel.

3. Free Radical and Radiation Biology Program, Carver College of Medicine, University of Iowa, Iowa 52242, United States.

4. School of Healthcare Science and Engineering, Vellore Institute of Technology, Vellore 632014, Tamil Nadu, India

5. Department of Inorganic Chemistry, Faculty of Natural Sciences, Comenius University Bratislava Ilkovičova 6, Mlynská dolina 842 15 Bratislava, Slovakia.

**Abstract**

In this work, we have studied the effect of base-pairing on the shape resonances of guanine and cytosine nucleobases. Among the seven π* resonances we identified in the guanine-cytosine (GC) anion radical, three were centered on cytosine, and the remaining were guanine-centered. Relative to the isolated bases, upon base pair formation, the cytosine resonances were red shifted, while the guanine-centered states showed an opposite trend - where their energy was blue shifted. In addition to the electronic interactions, geometric distortion and basis set superposition error plays a crucial role in the resonance positions and widths of the GC radical anion. The electronic interaction from the complementary base seems to have a larger effect on the stabilization of the anionic resonances than the surrounding environment.

*achintya@chem.iitb.ac.in

I. Introduction

Metastable anions of DNA play a major role in low-energy electron (LEE) induced radiation damage to genetic material.[1,2] Experimental works have suggested the involvement of transient negative ions (TNIs) in LEE induced DNA strand breaks.[3–5] Initial studies[3,6] highlighting the contribution of LEEs led to numerous experimental[7–15] and theoretical[16–26] studies to understand the mechanism involved in the process. TNIs have finite lifetimes and can decay by autodetachment of the extra electron, by dissociative electron attachment (DEA), or by relaxation to the ground state of a bound radical anion. For DNA strand breaks to occur, the TNIs must have a large enough lifetime for DEA to take place. During this period, the extra electron in the TNI can populate the σ*-type molecular orbital (MO) corresponding to the sugar-phosphate or sugar-nucleobase bonds, leading to their subsequent cleavage. Consequently, DEA in genetic material can result in single-strand breaks, double-strand breaks, and clustered damages.[1,2,27–31] In practice, TNIs are metastable states, called resonances, and they are broadly classified into two categories: shape type resonances and core-excited (Feshbach) resonances. Shape type resonances arise when the incoming electron is temporarily trapped behind a molecular potential barrier.[32] Core-excited resonances involve attachment and excitation simultaneously, i.e. the incoming electron attaches to an electronically excited state of the molecule. In this work, we focus our efforts on the shape type resonances of DNA.

Due to their short lifetimes (on the order of $\sim 10^{-15} - 10^{-12}$ sec),[32] experimental analysis of metastable states is challenging. Nevertheless, *in-silico* modeling of resonances is also challenging, since they are decaying (non-stationary) states. As such, they cannot be represented by a single discrete eigenfunction of the time independent Schrödinger equation (TISE).[32] Hence, conventional electron structure methods used for bound-state calculations are not suitable for calculating resonances. However, within non-Hermitian quantum mechanics, these states can be described as a single eigenstate, where the eigenvalue is complex, and takes on the form: $E = E_r - i\Gamma/2$, in which, $E_r$ is the resonance position, and $\Gamma$ is the width.[32] The resonance width is related to the lifetime as follows: $\tau = \hbar/\Gamma$. Within this formalism, several theoretical approaches have been described in the literature, including complex absorption potential (CAP),[33–36] complex basis functions,[37–39] complex scaling[40] and more. A major drawback of these approaches is that they require significant modifications to the existing bound-state quantum chemistry codes.

Alternatively, there are approaches for simulating resonances that apply conventional, Hermitian, quantum mechanical methods without any modifications.[41–46] These are stabilization-based methods, and they involve constructing a stabilization plot.[44–46] This plot describes the energy spectrum of the molecule as a function of a non-physical, numerical parameter. The data obtained from the stabilization plot is then analytically continued into the complex energy plane to obtain the complex eigenvalue of the resonance state.

Two popular stabilization-based methods in the literature are the general Padé approximation (GPA) and the Resonance Via Padé (RVP) technique.[41–43,47] The key difference between the two lies in the type of data from the stabilization curve they use, and in how they perform analytic continuation into the complex energy plane.[42] In this study, we use the RVP method to determine the resonance parameters (position and width) of anionic resonances. The RVP method, introduced by Moiseyev and co-workers, has been extensively employed to simulate the resonances of atomic systems (e.g. $Be^*$, $H^{*-}$, $He^*$) and small molecules like uracil, as well as microsolvated cytosine.[41,47–49] It can also be used to determine properties such as complex transition dipoles between TNIs.[48] The computational bottleneck in the stabilization-based approaches is the construction of the stabilization plot itself. It is crucial to employ accurate electronic structure methods together with sufficiently large, diffuse-augmented basis sets to obtain this graph. The equation-of-motion coupled-cluster method for electron attachment at the singles and doubles level (EA-EOM-CCSD) is an ideal candidate for this purpose. The recent domain-based local pair natural orbital implementation (EA-EOM-DLPNO-CCSD)[50] enables applications to moderately large systems. Hence, we use RVP combined with EA-EOM-DLPNO-CCSD (RVP-EA-EOM-DLPNO-CCSD) to simulate TNIs in DNA.

Computational analyses of LEE attachment to genetic material typically use small DNA model systems, because the large size of DNA makes quantum mechanical simulations on entire strands practically impossible. Purine and pyrimidine nucleobases, base pairs, and mono- and diphosphate nucleotides are commonly employed in the literature to model DNA.[18–20,22–26,51–57] However, theoretical investigations into anionic resonances in DNA model systems have predominantly focused on nucleobases.[44,45,48,49] The surrounding medium of DNA can also influence TNIs, since electron attachment to the genetic material occurs within the cellular medium. Recently, water microsolvated nucleobases were used to model TNIs of DNA in aqueous environments,[49,58] and the effect of amino acid environments on the nucleobase centered resonances has also been studied.[59] Nevertheless, nucleobase-centered DNA resonances are more likely to be strongly influenced by the complementary base than by the solvent. The interactions between the complementary nucleobases are stronger than those between a nucleobase and surrounding solvent molecules. In addition to the strong hydrogen-bonding, inter-base charge transfer can also affect the electron attachment process. Therefore, the Watson-Crick guanine-cytosine (GC) base pair is chosen as the model system to study anionic resonance states in genetic material.

**II. Computational Details**

Geometries of the GC base pair, cytosine, and guanine were optimized at RI-MP2/def2-TZVP level of theory. Subsequent vibrational frequency analyses were also conducted to ensure the geometries were at a local minimum of their respective potential energy surfaces. The energy spectrum of these systems was calculated using the EA-EOM-DLPNO-CCSD[50] method with the NORMALPNO setting and the cc-pVDZ(+2s2p2d) basis set. This choice follows Bouskila et al. analysis on the impact of various basis sets from 6-31+G to cc-pVTZ(+2s2p2d) on the resonance width and position of isolated uracil.[48] Bouskila et al. proposed cc-pVTZ(+2s2p2d) as the optimal basis set for modeling nucleobase shape resonances. However, the GC base pair is a larger system than pyrimidine nucleobases. Consequently, given that constructing a stabilization plot requires hundreds of EA-EOM-DLPNO-CCSD calculations, we selected Dunning's cc-pVDZ basis set with augmented diffuse functions to balance cost and accuracy. The cc-pVDZ(+2s2p2d) basis set was constructed by adding two extra diffuse functions for each of the s, p, and d angular momenta in an even-tempered manner to cc-pVDZ. Each extra diffuse Gaussian function was generated by dividing the exponent of the most diffuse function of the same angular momentum by a factor of 2.0. The stabilization plot for these systems was generated by scaling the exponents of the added diffuse functions by the parameter $\alpha$, i.e. dividing the exponents by $\alpha$, where $\alpha$ varied from 0.1 to 1.8. The stable regions that correspond to the resonance states where identified from the stabilization plots.[41] We then plotted the natural orbitals at several points within the stable region to confirm that they correspond to the same resonance state. The stabilization plot for the GC base pair is shown in Fig. 1, with the stable regions highlighted. The resonance parameters were then calculated based on the stable regions using the RVP method implemented in the open-source "Automatic RVP" software.[60] All calculations were performed using the ORCA software package (version 5.0.3).[61,62]

## III. Results and Discussions

### III.A Shape resonance of isolated cytosine and Guanine:

The primary objective of this work is to understand the impact of base pairing on shape resonances of isolated nucleobases. We have chosen GC base pair and the effect of base-pairing on the constituent nucleobases of GC is then analyzed by comparing their resonance parameters with those of GC base pair anion. In this work, the cytosine $1\pi^*$ state will be represented as C-$1\pi^*$ and that of guanine nucleobase as G-$1\pi^*$. Following the same convention, the first shape resonance of the GC base pair is called GC-$1\pi^*$.

Shape resonances of cytosine nucleobase have been extensively studied in the literature.[35,45,49,63–66] Recent work from our group has analyzed the effect of microsolvation on the shape resonances of cytosine at the RVP-EA-EOM-DLPNO-CCSD/aug-cc-pVDZ(+1s1p1d) level of theory.[49] In the present work, we were able to identify three low-lying cytosine resonances, as depicted in Fig. 2(a). Results from the RVP calculation showed that the $1\pi^*$, $2\pi^*$, and $3\pi^*$-type resonances of cytosine have energies of 0.73, 2.22, and 5.28 eV, respectively (See Table 1). As expected, the lowest shape resonance is the long-lived state with a calculated lifetime of 109 fs. The lifetime decreases drastically when one moves to $2\pi^*$ (31 fs) and $3\pi^*$ (6 fs) resonances.

Table 1 presents our results alongside representative theoretical and experimental data. Relative to Verma et al.,[49] with RVP-EA-EOM-DLPNO-CCSD/aug-cc-pVDZ+1s1p1d, the energies of the first, second, and third states are redshifted by 0.17, 0.11, and 0.54 eV, respectively. In contrast, the position and width of first and second resonance states in elative aug-cc-pVQZ+1s1p1d basis set are broadly consistent with our results. This is consistent with the finding by Bouskila et al.[48] that the cc-pVXZ+2s2p2d hierarchy of basis sets, which were used in our work, are more suitable in describing anionic resonances using stabilization-based approaches than its augmented counterpart. Further examining Table 1, we can see that our resonance energies are red-shifted with respect to GPA-based EA-EOM-CCSD values,[45] and are consistent with SAC-CI(CAP) results.[35] R-matrix scattering calculations, by contrast, yield widely varying resonance positions and widths for cytosine.[61,63] Experimental gas-phase positions reported by Burrow and co-workers [65] are lower than all theoretical values in Table 1, although the spacing between peaks is similar to that obtained in this study.

Isolated Guanine base exists in both keto and enol tautomeric forms. The GC base pair, however, is constituted by the keto tautomer of guanine.[68] Therefore, we have only considered keto guanine in this work. Electron transmission spectroscopy (ETS) on gas-phase guanine revealed three low-lying shape resonances between 0.46 and 2.36 eV.[67] Burrow and co-workers ascertained that the experimental values correspond to the enol isomer based on theoretically determined vertical attachment energies (VAE).[67] However, Fennimore and Matsika argued that the calculations by Burrow and co-workers[67] are less reliable, and that the resonances actually correspond to the keto isomer.[45] Their studies using EA-EOM-CCSD/aug-cc-pVDZ level of theory combined with GPA, showed that the VEA of keto tautomer is lower than the enol tautomer of guanine.[45] Theoretical studies using R-matrix scattering calculations,[63,65] CAP/SAC-CI,[35] and the stabilization-based GPA[45] method have predicted four $\pi^*$ shape resonances for guanine thereby suggesting that the ETS experiments did not capture the high-lying resonance with energy >6.0 eV.[67] However, the stabilization plot we generated (Fig. S2) for guanine shows five shape resonances.

The natural orbital plots corresponding to the guanine resonances are represented in Fig. 2(b). Except for the $2\pi^*$ state of guanine, all of the remaining four resonance states were identified by Fennimore and Matsika in their GPA work.[45] They reported an energy position of 1.00 eV and a width of 0.74 eV for the lowest shape resonance of guanine, corresponding to a lifetime of 0.89 fs in the GPA-EA-EOM-CCSD method. This lifetime is significantly lower than that observed for other nucleobases in their study. In comparison, our calculated lifetime for this state is 51 fs, i.e. in a similar range to the result for cytosine (See Table 2).

The resonance energies determined using CAP/SAC-CI[35] also do not show a consistent trend compared to our RVP values: The $3\pi^*$ and $5\pi^*$ states have similar energies whereas that of the $1\pi^*$ resonance is redshifted at RVP-EA-EOM-DLPNO-CCSD level. At the same time, the $4\pi^*$ state position is blue shifted compared to CAP/SAC-CI level. At the same time, the widths we have obtained are blue shifted for all states compared to CAP/SAC-CI values. Similar to GPA, the G-$2\pi^*$ state was not identified in the studies by Sommerfeld and co-workers (CAP/SAC-CI),[35] and Sanchez and co-workers (SMC).[66] Other theoretical results obtained using scattering calculations did not provide the molecular orbitals corresponding to the guanine resonances in their work. Hence, one cannot ascertain whether G-$2\pi^*$ is present in their work or not. Since the second resonance of guanine determined by Matsika and co-workers corresponds to the G-$3\pi^*$ state in this work, we have compared the second resonances in the R-matrix-based[63,65] and SMC/SEP[64] scattering calculations with our G-3π* resonance (Table 2). For this state, our calculated width is significantly lower than that previously reported.

**III.B Effect of base-pairing:**

Theoretical efforts to model the TNIs of purine and pyrimidine bases have yielded widely varying results, with a large spread in resonance position and width.[35,45,49,63–66] As shown above, our RVP results fall within this spread, and tend to overestimate the resonance positions for cytosine and guanine relative to their experimental values. Since the sensitivity of electronic structure methods to the basis set can be a major factor determining the accuracy of the calculated resonance positions and widths,[49] we do not expect the cc-pVDZ(+2s2p2d) basis set to provide quantitatively correct results for the systems considered in this work. Nevertheless, our calculations portray the qualitative correlation between the resonances of the GC base pair and those of the individual nucleobases.

To the best of our knowledge, experimental measurements of resonances in Watson-Crick base pairs have not been reported. Moreover, to the best of our knowledge only one theoretical study on the position and width of TNIs of nucleobase pairs exists in the literature.[66] This could be attributed to the moderately large size of the base pairs, which makes any accurate computational calculations prohibitively expensive. Sanchez and co-workers studied the shape resonances of GC base pair using elastic scattering calculations under the static exchange approximation.[66] They identified six π*-type resonances for the system with energies ranging from 2.0 eV to 10.3 eV. Among the four lowest quasi-bound anionic states described in their work, two were guanine-centered, and two were cytosine-centered. The two cytosine-centered shape resonances in the GC pair were lower in energy and width compared to isolated gas-phase cytosine nucleobase. However, the opposite trend was observed for guanine, as the two low-lying guanine-centered resonances got destabilized in the presence of cytosine. Sanchez and co-workers attributed this observation to the charge transfer between the nucleobases.[66] They claimed that the cytosine to guanine charge transfer causes the cytosine-centered resonances to become stable and destabilizes the guanine-centered TNIs.

In our work, we identified seven π* shape resonances of the GC base pair from its stabilization plot (Fig. 1), with positions ranging from 0.45 eV (GC-1π*) to 7.03 eV (GC-7π*). Three of these GC resonances were cytosine-centered, and the remaining four were guanine-centered (see Fig. 3). The two lowest GC resonances correspond to the C-1π* and C-2π* states. Whereas the GC resonance corresponding to the C-3π* state appears as the sixth state (GC-6π*) rather than the third. The guanine-centered GC states are GC-3π* to GC-5π*, with an additional guanine-centered state at GC-7π *.

Relative to isolated cytosine, the cytosine-centered resonances in the GC are stabilized and longer lived. The resonance position of C-1π* redshifts by 0.28 eV upon pairing with guanine (GC-1π*). The third resonance of cytosine also redshifts by a similar extent, from 5.28 eV (C-3π*) to 4.95 eV (GC-6π*). The largest redshift is observed for C-2π*, whose energy decreases from 2.59 eV in cytosine to 1.60 eV in the GC base pair. At the sametime the C-1π* lifetime increases dramatically from 109 fs to 213 fs (GC-1π*), whereas the C-2π* and C-3π* lifetimes increase by only 11 fs and 1 fs, respectively, in the GC resonances (GC-2π* and GC-6π*).

Comparing the resonances of guanine with those of the GC, all guanine resonances except for G-1π* were observed in the GC base pair. As shown in Fig. 3, G-2π*, G-3π*, and G-4π* are similar to GC-3π*, GC-4π*, and GC-5π* respectively, and the high-lying fifth state of guanine (G-5π*) corresponds to the highest shape resonance of the GC (GC-7π*). Contrary to the stabilization effect observed for the cytosine resonances upon base pair formation, the GC resonances localized on guanine were blue-shifted compared to their values in isolated guanine. This observation is consistent with Sanchez and co-workers' results.[66] In their work, the two guanine-centered GC resonances correspond to GC-3π* and GC-4π* in the present study, although large deviations are present in the reported resonance positions and widths when compared to our results.

Having discussed cytosine- and guanine-centered resonances separately, we now examine the GC base pair as a coupled system. The GC resonances shown in Fig. S5 appear to be localized on cytosine or guanine also have contribution from the complementary nucleobase. GC-3π* originates from the mixing of C-1π* and G-2π*, with a larger contribution from the latter (Fig. S5). The GC-2π* resonance is dominated by C-2π*, with a smaller but distinct G-2π*-like character. GC-4π* has a minor contribution from C-2π*, although it is predominantly a guanine-localized state (as in G-3π*). GC-1π* closely resembles C-1π* but also consists of contribution from a guanine-centered state which is different from the rest five resonances of isolated guanine. When the high-lying resonances are considered, the guanine-centered GC-5π* and GC-7π* resonances include contributions from cytosine. However, the nature of the states does not correspond to the three cytosine shape resonances. Conversely, the cytosine-centered GC-6π* resonance involves contribution from guanine-centered state that differs from the resonances of isolated guanine. The delocalized nature of resonances in base-pair consistent with experimental results of Namaan and co-workers.[69]

**III.C The effect of basis set superposition error and geometric distortion:**

To elucidate the interaction between the two bases in GC, we evaluated the geometric distortion effect and basis set superposition error (BSSE) associated with base pairing. The geometric distortion effect represents energy change associated with altering the neutral equilibrium geometry of the isolated nucleobase to that within the base pair. It may destabilize the nucleobase resonance states.

BSSE artificially lowers the energy by increasing the basis set dimension of a molecule in a complex, resulting from overlap with the basis functions of the neighboring molecules. In a recent study from our group, the effect of microsolvation on the shape resonances of cytosine was analyzed.[49] Calculations involving monohydrated cytosine where water atoms were kept as ghosts to assess the effect of the basis set on the resonance parameters. Verma et al.[49] found that the stabilization observed for the monohydrated cytosine TNIs was largely due to the finite basis set artifact. One can potentially overcome the impact of BSSE by utilizing sufficiently large basis sets in computational calculations. However, in the case of GC, this approach is not practically feasible due to the large size of the system being analyzed.

To understand the effect of finite basis set artifact and geometric distortion effect, we performed two sets of calculations. In the first one, GC resonance parameters were determined by keeping the cytosine or guanine atoms as ghosts. In the second one, RVP-EA-EOM-DLPNO-CCSD calculations were performed on guanine and cytosine at their base pair geometry. GC with cytosine as ghost (GC-Cghost) has five $\pi^*$-type resonances (Fig. S6(b)), which are identical to those formed when vertical electron attachment occurs to guanine in equilibrium neutral geometry. A comparison of the resonances between these two systems would focus solely on the BSSE effect on the resonance parameters. As shown in Table 3, the shape resonances of GC-Cghost have lower energy and width as compared to the corresponding states for Guanine in GC geometry. We have performed similar calculations by keeping Guanine as ghosts (GC-Gghost) and for cytosine in GC geometry. Both systems exhibit three $\pi^*$-type TNIs (Fig. S6(a)) similar to isolated cytosine. The larger basis set dimension of GC-Gghost results in stabilization of its $\pi^*$ states compared to cytosine in GC geometry. This clearly establishes the stabilizing influence of BSSE on the nucleobase resonances in GC base pair.

To understand the effect of geometric distortion on the shape resonances, one could compare the resonance energy and width of a nucleobase in its equilibrium neutral geometry with those in the base-pair geometry. Fig. S10 highlights geometric parameters (bond length, bond angle, etc.) that show the largest difference between isolated nucleobases and GC. For both nucleobases, geometric distortion results in destabilization of the resonances (Table 3).

The trend seen in GC resonances compared to its constituent bases may be due to the combination of multiple factors, which includes mixing of resonances, BSSE, and geometric distortions. Among these factors, the last two show similar trend for all the states. In contrast, the effect of the mixing of resonance states depends upon the nature of interaction of the complementary bases.

## IV. Conclusions

We have modeled the shape resonances of GC base pair using the RVP-EA-EOM-DLPNO-CCSD method. We identified seven shape resonances of GC, out of which three were localized on cytosine and the remaining four on guanine. We compared these GC states with the corresponding states of the individual nucleobases in their isolated geometry. The guanine-centered resonances in GC were blue-shifted compared to the corresponding states in isolated guanine. The opposite trend was observed for cytosine-centered resonances, where the GC resonances localized on cytosine were red-shifted compared to corresponding states in isolated cytosine. This is presumably due to the mixing between the resonances of guanine to those of cytosine within the pair. Our calculations also indicate that in addition to the interaction of the cytosine and guanine centered resonances, geometric distortion can also play a role in dictating the nature of the resonance state in DNA base strands. The electronic interaction from the complementary base seems to have a larger effect than the surrounding environment described in ref [49,58]. The base pairing leads to partial delocalization of the resonance state over both bases and it points out stronger binding of electrons to double stranded DNA than that observed in single stranded one. However, a definitive comparison requires more realistic model systems like dinucleoside phosphate duplex. Work is in progress towards that direction.

**Supplementary Material**

Stabilization plots, natural orbitals for resonances corresponding to GC (guanine as ghost), GC (cytosine as ghost), guanine in GC geometry, and cytosine in GC geometry, and cartesian coordinates of GC and isolated nucleobases are provided in the supplementary material.

**Acknowledgments**

The authors acknowledge the support from the IIT Bombay, Prime Minister's Research Fellowship for financial support. IIT Bombay super computational facility, and C-DAC Supercomputing resources (Param Smriti, Param Brahma) for computational time. A.A. acknowledges Université Paris-Saclay for the Visiting Professorship at the Institut de Chimie Physique, for support from the Free Radical and Radiation Biology Program, and for a travel grant from the Holden Comprehensive Cancer Center, at the University of Iowa. AKD acknowledges the research fellowship funded by the EU NextGenerationEU through the Recovery and Resilience Plan for Slovakia under project No. 09I03-03-V04-00117.

**References**

[1] E. Alizadeh, T.M. Orlando, and L. Sanche, "Biomolecular damage induced by ionizing radiation: The direct and indirect effects of low-energy electrons on DNA," Annu. Rev. Phys. Chem. **66**, 379–398 (2015).

[2] J. Narayanan S J, D. Tripathi, P. Verma, A. Adhikary, and A.K. Dutta, "Secondary Electron Attachment-Induced Radiation Damage to Genetic Materials," ACS Omega **8**(12), 10669–10689 (2023).

[3] B. Boudaïffa, P. Cloutier, D. Hunting, M.A. Huels, and L. Sanche, "Resonant formation of DNA strand breaks by low-energy (3 to 20 eV) electrons," Science **287**(5458), 1658–1660 (2000).

[4] I. Bald, E. Illenberger, and J. Kopyra, "Damage of DNA by Low Energy Electrons (< 3 eV)," J. Phys. Conf. Ser. **373**(1), 012008 (2012).

[5] K. Ebel, and I. Bald, "Low-Energy (5–20 eV) Electron-Induced Single and Double Strand Breaks in Well-Defined DNA Sequences," J. Phys. Chem. Lett. **13**(22), 4871–4876 (2022).

[6] J. Woldhuis, J.B. Verberne, M.V.M. Lafleur, J. Retèl, J. Blok, and H. Loman, "γ-Rays Inactivate ɸX174 DNA in Frozen Anoxic Solutions at −20°C Mainly by Reactions of Dry Electrons," Int. J. Radiat. Biol. Relat. Stud. Phys. Chem. Med. **46**(4), 329–330 (1984).

[7] M.A. Huels, B. Boudaïffa, P. Cloutier, D. Hunting, and L. Sanche, "Single, double, and multiple double strand breaks induced in DNA by 3-100 eV electrons," J. Am. Chem. Soc. **125**(15), 4467–4477 (2003).

[8] H. Abdoul-Carime, S. Gohlke, E. Fischbach, J. Scheike, and E. Illenberger, "Thymine excision from DNA by subexcitation electrons," Chem. Phys. Lett. **387**(4–6), 267–270 (2004).

[9] Y. Zheng, P. Cloutier, D.J. Hunting, L. Sanche, and J.R. Wagner, "Chemical basis of DNA sugar-phosphate cleavage by low-energy electrons," J. Am. Chem. Soc. **127**(47), 16592–16598 (2005).

[10] L. Sanche, "Low energy electron-driven damage in biomolecules," Eur. Phys. J. D **35**(2), 367–390 (2005).

[11] Y. Zheng, P. Cloutier, D.J. Hunting, J.R. Wagner, and L. Sanche, "Phosphodiester and N-glycosidic bond cleavage in DNA induced by 4-15 eV electrons," J. Chem. Phys. **124**(6), (2006).

[12] G. Khorsandgolchin, L. Sanche, P. Cloutier, and J.R. Wagner, "Strand Breaks Induced by Very Low Energy Electrons: Product Analysis and Mechanistic Insight into the Reaction with TpT," J. Am. Chem. Soc. **141**(26), 10315–10323 (2019).

[13] B. Kumari, A. Huwaidi, G. Robert, P. Cloutier, A.D. Bass, L. Sanche, and J.R. Wagner, "Shape Resonances in DNA: Nucleobase Release, Reduction, and Dideoxynucleoside Products Induced by 1.3 to 2.3 eV Electrons," J. Phys. Chem. B **126**(28), 5175–5184 (2022).

[14] Y. Gao, Y. Dong, X. Wang, W. Su, P. Cloutier, Y. Zheng, and L. Sanche, "Comparisons between the Direct and Indirect Effect of 1.5 keV X-rays and 0–30 eV Electrons on DNA: Base Lesions, Stand Breaks, Cross-Links, and Cluster Damages," J. Phys. Chem. B **128**(45), 11041–11053 (2024).

[15] Y. Dong, X. Huang, W. Zhang, Y. Shao, P. Cloutier, Y. Zheng, and L. Sanche, "Hyperthermal Reactions in DNA Triggered by 1–20 eV Electrons: Absolute Cross Sections for Crosslinks, Strand Breaks, Clustered Damages and Base Modifications," Int. J. Mol. Sci. **26**(9), (2025).

[16] R. Barrios, P. Skurski, and J. Simons, "Mechanism for damage to DNA by low-energy electrons," J. Phys. Chem. B **106**(33), 7991–7994 (2002).

[17] X. Pan, P. Cloutier, D. Hunting, and L. Sanche, "Dissociative Electron Attachment to DNA," Phys. Rev. Lett. **90**(20), 208102 (2003).

[18] J. Simons, "How do low-energy (0.1-2 eV) electrons cause DNA-strand breaks?," Acc. Chem. Res. **39**(10), 772–779 (2006).

[19] J. Gu, Y. Xie, and H.F. Schaefer, "Near 0 eV electrons attach to nucleotides," J. Am. Chem. Soc. **128**(4), 1250–1252 (2006).

[20] M. Smyth, and J. Kohanoff, "Excess electron interactions with solvated DNA nucleotides: Strand breaks possible at room temperature," J. Am. Chem. Soc. **134**(22), 9122–9125 (2012).

[21] R. Bhaskaran, and M. Sarma, "The role of the shape resonance state in low energy electron induced single strand break in 2′-deoxycytidine-5′-monophosphate," Phys. Chem. Chem. Phys. **17**(23), 15250–15257 (2015).

[22] R. Bhaskaran, and M. Sarma, "Low-Energy Electron Interaction with the Phosphate Group in DNA Molecule and the Characteristics of Single-Strand Break Pathways," J. Phys. Chem. A **119**(40), 10130–10136 (2015).

[23] M. McAllister, M. Smyth, B. Gu, G.A. Tribello, and J. Kohanoff, "Understanding the Interaction between Low-Energy Electrons and DNA Nucleotides in Aqueous Solution," J. Phys. Chem. Lett. **6**(15), 3091–3097 (2015).

[24] M. McAllister, N. Kazemigazestane, L.T. Henry, B. Gu, I. Fabrikant, G.A. Tribello, and J. Kohanoff, "Solvation Effects on Dissociative Electron Attachment to Thymine," J. Phys. Chem. B **123**(7), 1537–1544 (2019).

[25] D. Davis, and Y. Sajeev, "A hitherto unknown stability of DNA basepairs," Chem. Commun. **56**(93), 14625–14628 (2020).

[26] C.S. Anstöter, M. Dellostritto, M.L. Klein, and S. Matsika, "Modeling the Ultrafast Electron Attachment Dynamics of Solvated Uracil," J. Phys. Chem. A **125**(32), 6995–7003 (2021).

[27] C. von Sonntag, *Free-Radical-Induced DNA Damage and Its Repair: A Chemical Perspective*, 1st ed. (Springer Berlin, Heidelberg, Berlin, Heidelberg, 2006).

[28] A. Adhikary, D. Becker, and M.D. Sevilla, "Electron Spin Resonance of Radicals in Irradiated DNA," in *Appl. EPR Radiat. Res.*, edited by A. Lund and M. Shiotani, (Springer International Publishing, Cham, 2014), pp. 299–352.

[29] M. Dizdaroglu, "Oxidatively induced DNA damage and its repair in cancer," Mutat. Res. Mutat. Res. **763**, 212–245 (2015).

[30] M.D. Sevilla, D. Becker, A. Kumar, and A. Adhikary, "Gamma and ion-beam irradiation of DNA: Free radical mechanisms, electron effects, and radiation chemical track structure," Radiat. Phys. Chem. Biomol. Recent Dev. **128**, 60–74 (2016).

[31] A. Kumar, D. Becker, A. Adhikary, and M.D. Sevilla, "Reaction of Electrons with DNA: Radiation Damage to Radiosensitization," Int. J. Mol. Sci. **20**(16), (2019).

[32] N. Moiseyev, *Non-Hermitian Quantum Mechanics* (Cambridge University Press, 2011).

[33] R. Santra, and L.S. Cederbaum, "Non-Hermitian electronic theory and applications to clusters," Phys. Rep. **368**, 1–117 (2002).

[34] A. Ghosh, N. Vaval, and S. Pal, "Equation-of-motion coupled-cluster method for the study of shape resonance.," J. Chem. Phys. **136 23**, 234110 (2012).

[35] Y. Kanazawa, M. Ehara, and T. Sommerfeld, "Low-Lying π∗ Resonances of Standard and Rare DNA and RNA Bases Studied by the Projected CAP/SAC-CI Method," J. Phys. Chem. A **120**(9), 1545–1553 (2016).

[36] T.C. Jagau, K.B. Bravaya, and A.I. Krylov, "Extending Quantum Chemistry of Bound States to Electronic Resonances," Annu. Rev. Phys. Chem. **68**, 525–553 (2017).

[37] T.N. Rescigno, C.W. McCurdy, and A.E. Orel, "Extensions of the complex-coordinate method to the study of resonances in many-electron systems," Phys. Rev. A **17**, 1931–1938 (1978).

[38] M. Hernández Vera, and T.-C. Jagau, "Resolution-of-the-identity second-order Møller–Plesset perturbation theory with complex basis functions: Benchmark calculations and applications to strong-field ionization of polyacenes," J. Chem. Phys. **152**(17), 174103 (2020).

[39] A.F. White, E. Epifanovsky, C.W. McCurdy, and M. Head-Gordon, "Second order Møller-Plesset and coupled cluster singles and doubles methods with complex basis functions for resonances in electron-molecule scattering," J. Chem. Phys. **146**(23), 234107 (2017).

[40] N. Moiseyev, and C. Corcoran, "Autoionizing states of H2 and H2- using the complex-scaling method," Phys Rev A **20**(3), 814–817 (1979).

[41] A. Landau, I. Haritan, P.R. Kaprálová-Žd'ánská, and N. Moiseyev, "Atomic and Molecular Complex Resonances from Real Eigenvalues Using Standard (Hermitian) Electronic Structure Calculations," J. Phys. Chem. A **120**(19), 3098–3108 (2016).

[42] I. Haritan, and N. Moiseyev, "On the calculation of resonances by analytic continuation of eigenvalues from the stabilization graph," J. Chem. Phys. **147**(1), 014101 (2017).

[43] A. Landau, I. Haritan, and N. Moiseyev, "The RVP Method—From Real Ab-Initio Calculations to Complex Energies and Transition Dipoles," Front. Phys. **10**, (2022).

[44] M.A. Fennimore, and S. Matsika, "Core-excited and shape resonances of uracil," Phys. Chem. Chem. Phys. **18**(44), 30536–30545 (2016).

[45] M.A. Fennimore, and S. Matsika, "Electronic Resonances of Nucleobases Using Stabilization Methods," J. Phys. Chem. A **122**(16), 4048–4057 (2018).

[46] M.F. Falcetta, L.A. DiFalco, D.S. Ackerman, J.C. Barlow, and K.D. Jordan, "Assessment of Various Electronic Structure Methods for Characterizing Temporary Anion States: Application to the Ground State Anions of N2, C2H2, C2H4, and C6H6," J. Phys. Chem. A **118**(35), 7489–7497 (2014).

[47] A. Landau, and I. Haritan, "The Clusterization Technique: A Systematic Search for the Resonance Energies Obtained via Padé," J. Phys. Chem. A **123**(24), 5091–5105 (2019).

[48] G. Bouskila, A. Landau, I. Haritan, N. Moiseyev, and D. Bhattacharya, "Complex energies and transition dipoles for shape-type resonances of uracil anion from stabilization curves via Padé," J. Chem. Phys. **156**(19), 194101 (2022).

[49] P. Verma, M. Mukherjee, D. Bhattacharya, I. Haritan, and A.K. Dutta, "Shape resonance induced electron attachment to cytosine: The effect of aqueous media," J. Chem. Phys. **159**(21), 214303 (2023).

[50] A.K. Dutta, M. Saitow, B. Demoulin, F. Neese, and R. Izsák, "A domain-based local pair natural orbital implementation of the equation of motion coupled cluster method for electron attached states," J. Chem. Phys. **150**(16), 164123 (2019).

[51] D. Tripathi, and A.K. Dutta, "Bound anionic states of DNA and RNA nucleobases: An EOM-CCSD investigation," Int. J. Quantum Chem. **119**(9), e25875 (2019).

[52] D. Tripathi, and A.K. Dutta, "Electron Attachment to DNA Base Pairs: An Interplay of Dipole- and Valence-Bound States," J. Phys. Chem. A **123**(46), 10131–10138 (2019).

[53] S. Ranga, M. Mukherjee, and A.K. Dutta, "Interactions of Solvated Electrons with Nucleobases: The Effect of Base Pairing," ChemPhysChem **21**(10), 1019–1027 (2020).

[54] J. Narayanan S J, D. Tripathi, and A.K. Dutta, "Doorway Mechanism for Electron Attachment Induced DNA Strand Breaks," J. Phys. Chem. Lett. **12**(42), 10380–10387 (2021).

[55] J. Narayanan S J, A. Bachhar, D. Tripathi, and A.K. Dutta, "Electron Attachment to Wobble Base Pairs," J. Phys. Chem. A **127**(2), 457–467 (2023).

[56] A. Kumar, M.D. Sevilla, and L. Sanche, "How a Single 5 eV Electron Can Induce Double-Strand Breaks in DNA: A Time-Dependent Density Functional Theory Study," J. Phys. Chem. B **128**(17), 4053–4062 (2024).

[57] J. Narayanan S J, P. Verma, A. Adhikary, and A. Kumar Dutta, "Electron Attachment to the Nucleobase Uracil in Diethylene Glycol: The Signature of a Doorway," ChemPhysChem **25**(24), e202400581 (2024).

[58] J. Narayanan S J, D. Tripathi, I. Haritan, and A.K. Dutta, "The Effect of Aqueous Medium on Nucleobase Shape Resonances: Insights from Microsolvation," J. Phys. Chem. A **129**(48), 11179–11188 (2025).

[59] S. Arora, J. Narayanan S J, I. Haritan, A. Adhikary, and A.K. Dutta, "Effect of protein environment on the shape resonances of RNA pyrimidine nucleobases: Insights from a model system," J. Chem. Phys. **163**(13), 134103 (2025).

[60] Y. Safrai, and I. Haritan, "automatic-rvp: RVP Program," (2021).

[61] F. Neese, F. Wennmohs, U. Becker, and C. Riplinger, "The ORCA quantum chemistry program package," J. Chem. Phys. **152**(22), 224108 (2020).

[62] F. Neese, "Software update: The ORCA program system—Version 5.0," WIREs Comput. Mol. Sci. **12**(5), e1606 (2022).

[63] A. Dora, L. Bryjko, T. van Mourik, and J. Tennyson, "R-matrix study of elastic and inelastic electron collisions with cytosine and thymine," J. Phys. B At. Mol. Opt. Phys. **45**(17), 175203 (2012).

[64] C. Winstead, V. McKoy, and S. d'Almeida Sanchez, "Interaction of low-energy electrons with the pyrimidine bases and nucleosides of DNA.," J. Chem. Phys. **127**(8), 085105 (2007).

[65] S. Tonzani, and C.H. Greene, "Low-energy electron scattering from DNA and RNA bases: shape resonances and radiation damage.," J. Chem. Phys. **124**(5), 054312 (2006).

[66] F.B. Nunes, M.T. do Nascimento Varella, D.F. Pastega, T.C. Freitas, M.A.P. Lima, M.H.F. Bettega, and S. d'Almeida Sanchez, "Transient negative ion spectrum of the cytosine-guanine pair," Eur. Phys. J. D **71**(4), 92 (2017).

[67] K. Aflatooni, G.A. Gallup, and P.D. Burrow, "Electron Attachment Energies of the DNA Bases," J. Phys. Chem. A **102**(31), 6205–6207 (1998).

[68] V.A. Bloomfield, and D.M. Crothers, "Nucleic acids: structures, properties and functions. By Victor A. Bloomfield, Donald M. Crothers and Ignacio Tinoco Jr. Pp. xii + 794. Sausalito: University Science Books, 2000. Price US$79.20. ISBN 0-935702-49-0.," Acta Crystallogr. Sect. B-Struct. Sci. **57**, 602–602 (2001).

[69] S.G. Ray, S.S. Daube, and R. Naaman, "On the capturing of low-energy electrons by DNA," Proc. Natl. Acad. Sci. **102**(1), 15–19 (2005).

Table 1. Comparison of resonance positions and widths (in parentheses) for isolated cytosine determined in the present study with previous theoretical and experimental works. Values are given in eV.

| Method | $1\pi^*$ | $2\pi^*$ | $3\pi^*$ |
| --- | --- | --- | --- |
| RVP-EA-EOM-DLPNO-CCSD/cc-pVDZ+2s2p2d[a] | 0.73 (0.006) | 2.22 (0.022) | 5.28 (0.109) |
| RVP-EA-EOM-DLPNO-CCSD/aug-cc-pVDZ+1s1p1d (NORMALPNO)[49] | 0.90 (0.0035) | 2.33 (0.03) | 5.82 (0.183) |
| RVP-EA-EOM-DLPNO-CCSD/aug-cc-pVQZ+1s1p1d (NORMALPNO)[49] | 0.70 (0.002) | 2.20 (0.021) | 4.72 (0.133) |
| GPA-EA-EOM-CCSD/aug-cc-pVDZ[45] | 0.93 (0.017) | 2.40 (0.19) | 5.54 (0.35) |
| CAP/SAC-CI[35] | 0.70 (0.16) | 2.18 (0.30) | 5.66 (0.63) |
| R-matrix/u-CC[63] | 0.36 (0.016) | 2.05 (0.30) | 5.35 |
| R-matrix/SEP[63] | 0.71 (0.05) | 2.66 (0.33) | 6.29 (0.72) |
| SMC/SEP[64] | 0.50 | 2.40 | 6.30 |
| R-matrix/SE[65] | 1.70 (0.50) | 4.30 (0.70) | 8.10 (0.80) |
| SMC[66] | 0.61 (0.24) | 1.74 (0.66) | 5.50 |
| expt[67] | 0.32 | 1.53 | 4.50 |

a. Current work

Table 2. Comparison of resonance positions and widths (in parentheses) for isolated guanine determined in the present study with previous theoretical and experimental works. Values are given in eV.

| Method | 1π* | 2π* | 3π* | 4π* | 5π* |
| --- | --- | --- | --- | --- | --- |
| RVP-EA-EOM-DLPNO-CCSD/cc-pVDZ+2s2p2d[a] | 1.39 (0.013) | 1.48 (0.021) | 1.68 (0.026) | 2.75 (0.030) | 6.51 (0.060) |
| GPA-EA-EOM-CCSD/aug-cc-pVDZ[45] | 1.00 (0.74) | | 1.70 (0.10) | 3.06 (0.09) | 6.64 (0.21) |
| CAP/SAC-CI[35] | 1.11 (0.64) | | 1.66 (0.34) | 3.21 (0.55) | 6.50 (0.92) |
| R-matrix/u-CC[63] | 0.97 (0.006) | | 2.41 (0.15) | 3.78 (0.29) | 6.42 (0.72) |
| R-matrix/SEP[63] | 1.83 (0.16) | | 3.30 (0.24) | 4.25 (0.33) | 7.36 (0.27) |
| SMC/SEP[64] | 1.55 | | 2.40 | 3.75 | |
| R-matrix/SE[65] | 2.00 (0.20) | | 3.80 (0.25) | 4.80 (0.35) | 8.90 (0.60) |
| SMC[66] | 0.90 (0.51) | | 1.55 (0.70) | 2.75 | |
| expt[67] | 0.46 | | 1.37 | 2.36 | |

a. Current work

Table 3. The resonance positions and widths (in parentheses) of all the systems considered in this work calculated with the RVP-EA-EOM-DLPNO-CCSD method and cc-pVDZ(+2s2p2d) basis set.

| System | 1π* | 2π* | 3π* | 4π* | 5π* | 6π* | 7π* |
|---|---|---|---|---|---|---|---|
| GC base pair | 0.45 (0.003) | 1.60 (0.022) | 1.86 (0.060) | 2.47 (0.077) | 3.65 (0.102) | 4.95 (0.096) | 7.03 (0.088) |
| Guanine | 1.39 (0.013) | 1.48 (0.021) | 1.68 (0.026) | 2.75 (0.030) | 6.51 (0.060) | | |
| Cytosine | 0.73 (0.006) | 2.22 (0.022) | 5.28 (0.109) | | | | |
| GC (guanine as ghost) | 0.81 (0.009) | 2.33 (0.036) | 5.68 (0.124) | | | | |
| GC (cytosine as ghost) | 1.49 (0.015) | 1.59 (0.025) | 1.80 (0.029) | 3.15 (0.034) | 6.74 (0.075) | | |
| Guanine in GC geometry | 1.60 (0.014) | 1.60 (0.025) | 1.88 (0.034) | 3.16 (0.048) | 7.20 (0.140) | | |
| Cytosine in GC geometry | 0.82 (0.008) | 2.41 (0.061) | 5.82 (0.123) | | | | |

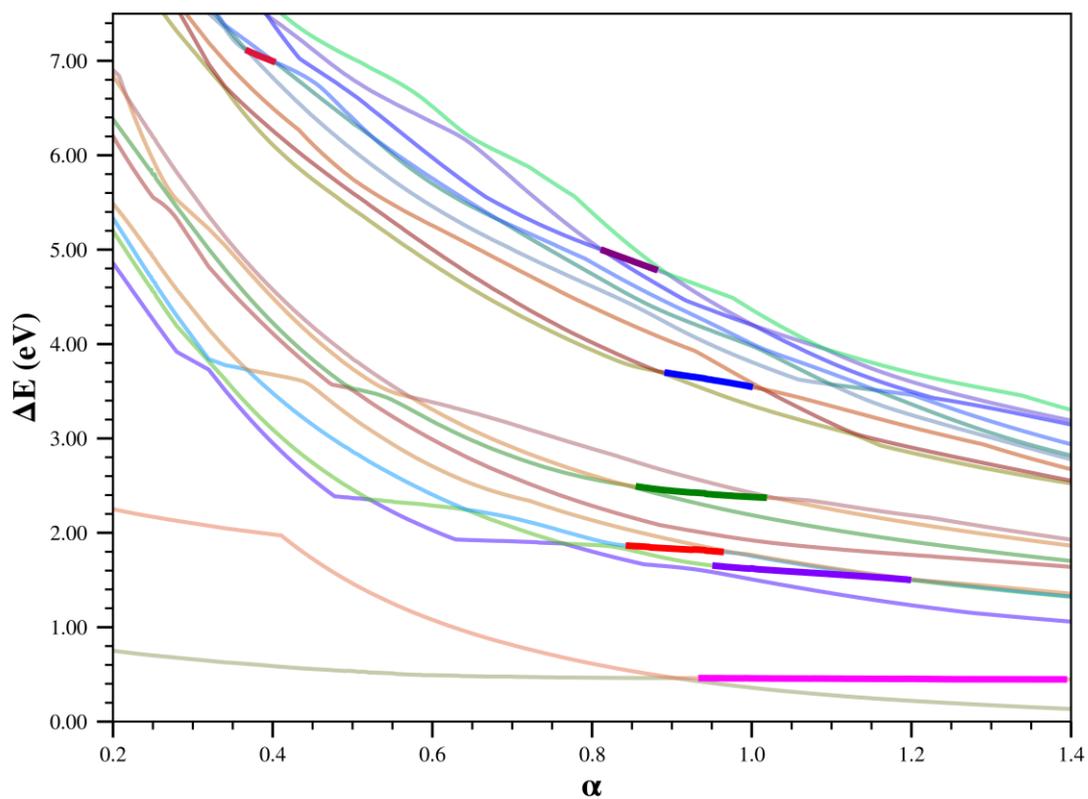

Fig. 1. Stabilization plot for GC base pair anion generated at EA-EOM-DLPNO-CCSD level of theory with cc-pVDZ(+2s2p2d) basis set. The stable regions of the seven resonances are highlighted.

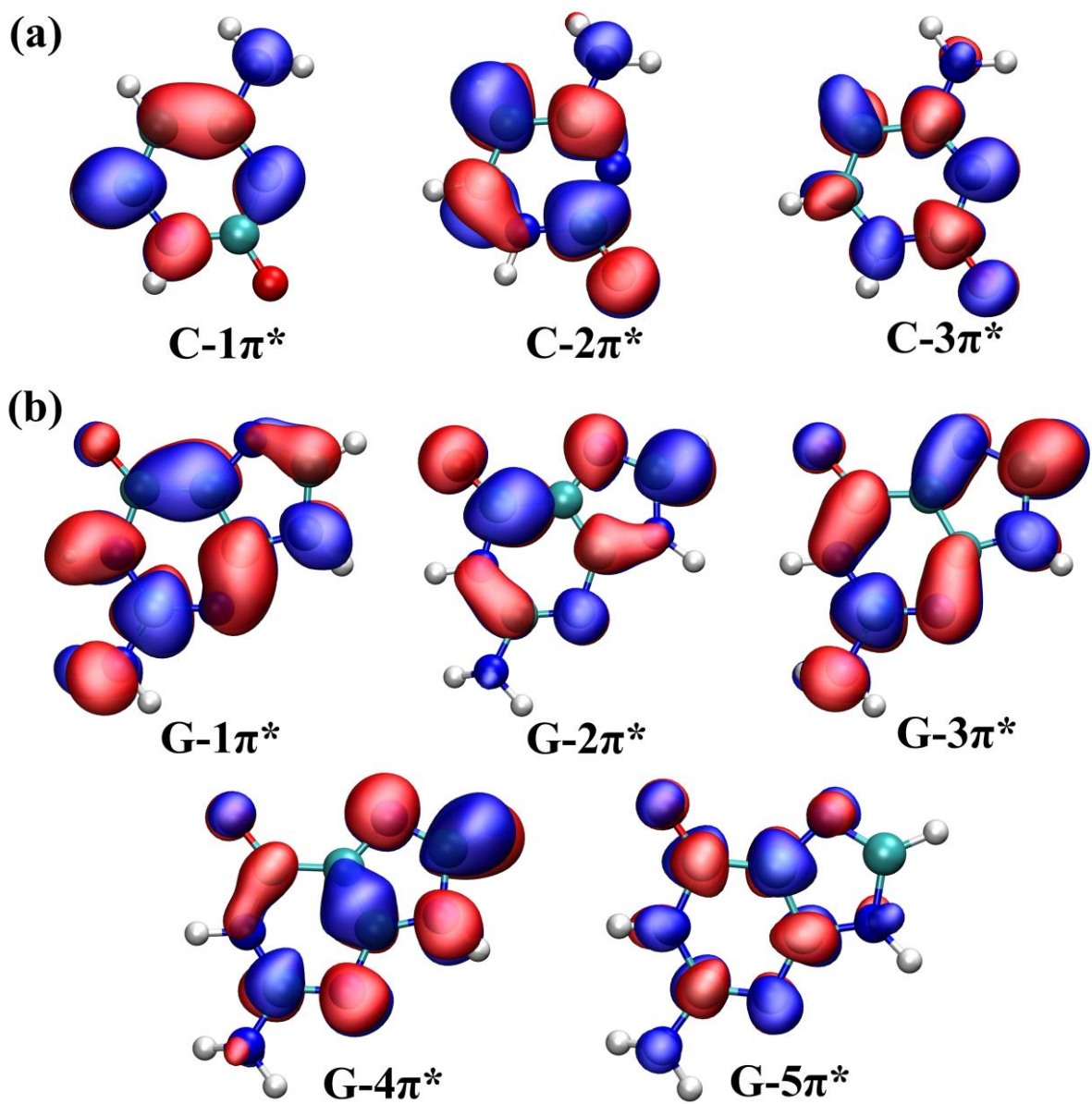

Fig. 2. The EA-EOM-DLPNO-CCSD/cc-pVDZ(+2s2p2d) natural orbitals corresponding to the π* shape resonances of (a) cytosine (b) guanine in their equilibrium geometries.

| System | 1π* | 2π* | 3π* | 4π* | 5π* | 6π* | 7π* |
|---|---|---|---|---|---|---|---|
| GC base pair | **0.45 (0.003)** | **1.60 (0.022)** | **1.86 (0.060)** | **2.47 (0.077)** | **3.65 (0.102)** | **4.95 (0.096)** | **7.03 (0.088)** |
| Guanine | 1.39 (0.013) | **1.48 (0.021)** | **1.68 (0.026)** | **2.75 (0.030)** | **6.51 (0.060)** | | |
| Cytosine | **0.73 (0.006)** | **2.22 (0.022)** | **5.28 (0.109)** | | | | |

**GC and Guanine**

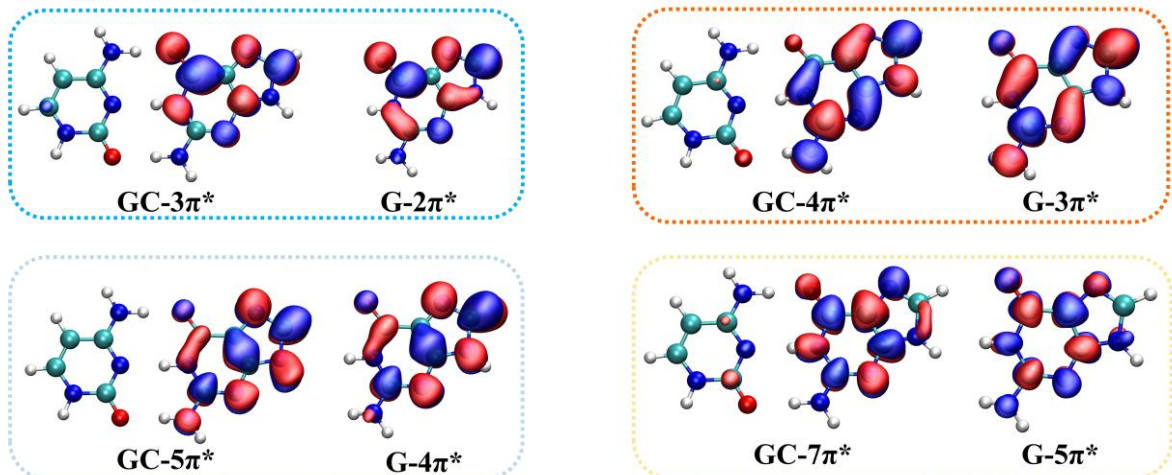

**GC and Cytosine**

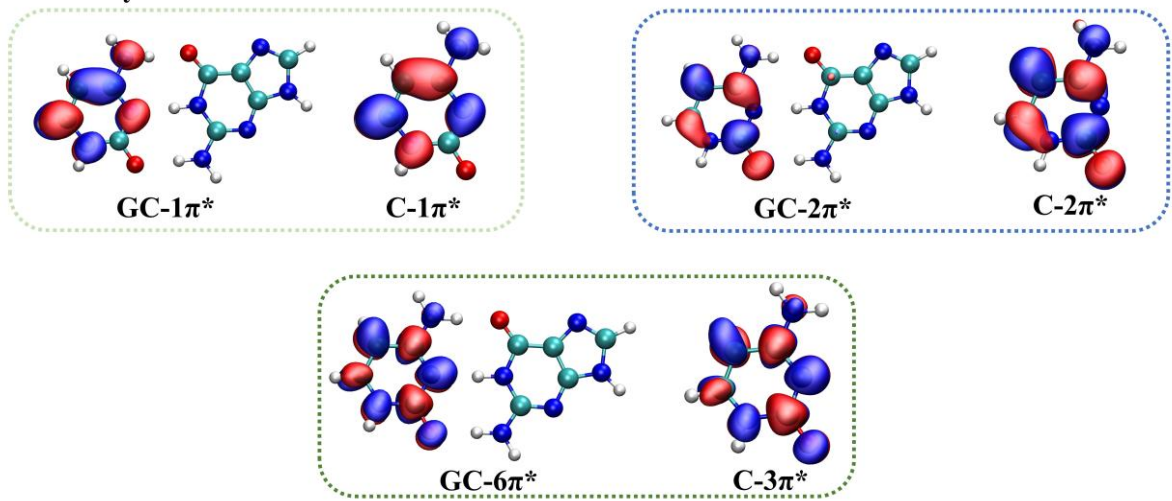

Fig. 3. The table consists of resonance positions and widths (in parentheses) of the GC base pair, guanine, and cytosine calculated with the RVP-EA-EOM-DLPNO-CCSD method and cc-pVDZ(+2s2p2d) basis set. Resonance states in GC correspond to those in the isolated nucleobases are highlighted using the same color. The corresponding MO plots of the GC and nucleobases resonances are shown below the table. Values are given in eV.